\documentclass[twocolumn,prd]{revtex4-1}

\usepackage{graphicx}
\usepackage{amssymb}
\usepackage{amsmath}

\newcommand{\abs}[1]{|#1|}

\date{\today}

\begin{document}

\author{Kristjan Kannike}
 \email{Kristjan.Kannike@cern.ch}
\affiliation{National Institute of Chemical Physics and Biophysics, \\
R\"{a}vala 10, Tallinn 10143, Estonia}
\author{Martti Raidal}%
 \email{Martti.Raidal@cern.ch}
\affiliation{National Institute of Chemical Physics and Biophysics, \\
R\"{a}vala 10, Tallinn 10143, Estonia}

\title{Phase transitions and gravitational wave tests of pseudo-Goldstone dark matter in the softly broken U(1) scalar singlet model}

\begin{abstract}
 We study phase transitions in a softly broken $U(1)$ complex singlet scalar model in which the dark matter is the pseudo-scalar part of a singlet whose direct detection coupling to matter is strongly suppressed. Our aim is to find ways to test this model with the stochastic gravitational wave background from the scalar phase transition. We find that the phase transition which induces vacuum expectation values for both the Higgs boson and the singlet -- necessary to provide a realistic dark matter candidate -- is always of the second order. If the stochastic gravitational wave background characteristic to a first order phase transition will be discovered by interferometers, the soft breaking of $U(1)$ cannot be the explanation to the suppressed dark matter-baryon coupling, providing a conclusive negative test for this class of singlet models.
\end{abstract}

\maketitle

\section{Introduction}


Scalar singlet is one of the most generic candidates for the dark matter (DM) of Universe~\cite{Silveira:1985rk,McDonald:1993ex}, 
whose properties have been exhaustively studied~\cite{Barger:2008jx,Burgess:2000yq,Cline:2013gha,Djouadi:2011aa} (see~\cite{Athron:2017kgt,Arcadi:2019lka} for a recent review and references). However, the recent results from direct detection experiments~\cite{Akerib:2016vxi,Aprile:2018dbl,Cui:2017nnn} have pushed the singlet scalar DM mass above a TeV-scale (except in a narrow region around the Higgs resonance). Thus, the singlet scalar models with the simplest scalar potential, in which the DM is stabilised by a $\mathbb{Z}_{2}$ symmetry, appear to be strongly constrained, less natural and less attractive.

This conclusion need not hold for specific realisations of the singlet scalar DM idea. A neat observation was made in~\cite{Gross:2017dan} that for the case of a
less general scalar potential obtained by imposing an $U(1)$ symmetry that is softly broken, first studied in~\cite{Chiang:2017nmu}, the direct detection cross section is strongly suppressed at tree level by the destructive interference between two contributing amplitudes. This result persists even if loop-level corrections to the direct detection cross section are 
considered~\cite{Azevedo:2018exj,Ishiwata:2018sdi}, making the softly broken scalar singlet model really interesting. This has motivated follow-up studies demonstrating that it is possible for pseudo-Goldstone DM to show up at the LHC~\cite{Huitu:2018gbc} or in indirect detection~\cite{Alanne:2018zjm}.

Is there any other way to test the softly broken $U(1)$ singlet DM model experimentally and to distinguish the particular model from more general versions of singlet scalar DM? A new probe of physics beyond the Standard Model (SM) became experimentally available due to the discovery of gravitational waves (GWs) by LIGO experiment~\cite{Abbott:2016blz,Abbott:2016nmj}. It is well known that first-order phase transitions generate a stochastic GW background~\cite{Hogan:1984hx,Steinhardt:1981ct,Witten:1984rs}, which can potentially be probed in future space based GW interferometers~\cite{Corbin:2005ny,Seoane:2013qna}.
While the Higgs phase transition in the SM is of second order~\cite{Kajantie:1996mn,Aoki:1999fi} and, thus, does not generate the GW signal, 
in models with extended scalar sector the first-order phase transition in the early Universe can become experimentally testable by the GW experiments. (For a recent review on phase transitions and GWs, see \cite{Mazumdar:2018dfl}.)

GWs from the extension of the SM with a scalar singlet have been extensively studied. In general a two-step phase transition will take place in those models that can be of the first order \cite{Espinosa:2011eu,Cline:2012hg,Espinosa:2011ax,Alanne:2014bra,Alanne:2016wtx,Tenkanen:2016idg,Zhou:2018zli,Cheng:2018ajh} and be testable with GWs \cite{Kakizaki:2015wua,Hashino:2016xoj,Vaskonen:2016yiu,Huang:2016cjm,Artymowski:2016tme,Beniwal:2017eik,Chao:2017vrq,Bian:2017wfv,Huang:2018aja,Beniwal:2018hyi,Croon:2018erz}. The aim of this work is to study the properties of the phase transition in the scalar singlet model with a softly broken $U(1)$ symmetry in order to find out whether the GW signal can distinguish between different versions of the singlet DM models. We reach a definitive conclusion: in this class of models with a suppressed direct detection cross section, the phase transition is necessarily of the second order and no testable GW background will be generated. Therefore, if the stochastic GW background characteristic to the first order phase transition due to scalar singlets will be discovered, the softly broken singlet model cannot be responsible for that. In this case, as a consequence, the negative results from DM direct detection experiments cannot be explained with the ideas presented in~\cite{Gross:2017dan}. On the other hand, note that not discovering a GW signal would not rule out models with first-order phase transitions, because to generate a large signal, the phase transition must be strongly first-order.

This Letter is organised as follows. We describe the model in Section~\ref{sec:model}. The phase transition in this framework is studied in Section~\ref{sec:phase}. We conclude in Section~\ref{sec:concl}.

\vspace{4cm}

\begin{figure*}[tb]
\begin{center}
  \includegraphics{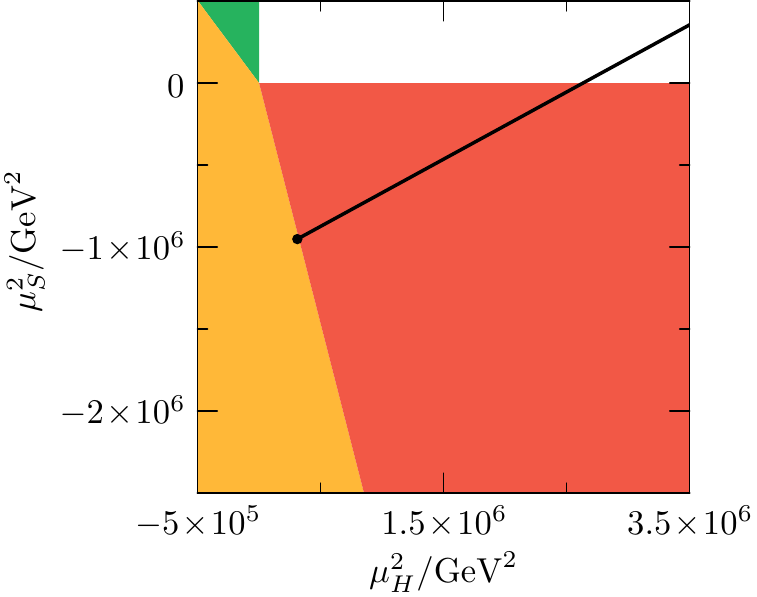}~~~~~
  \includegraphics{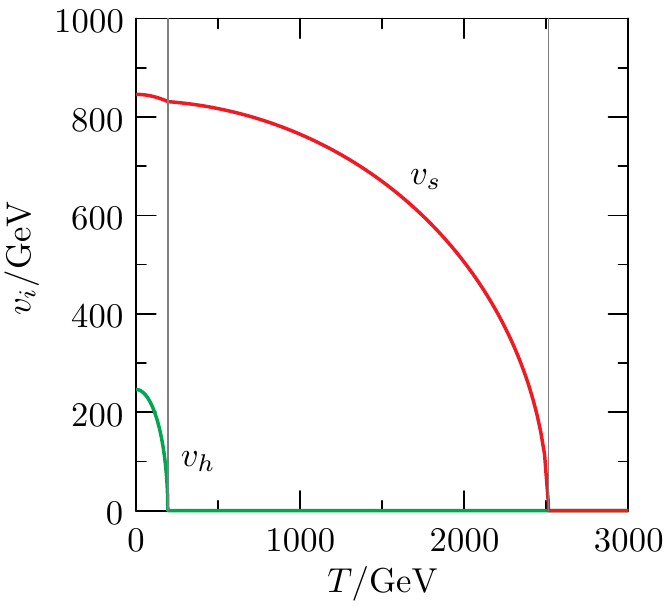}
\caption{Phase diagram and thermal evolution of VEVs in the considered model. \textit{Left panel:} Thermal evolution of the field to $T = 0$ (dot) is shown by the black line. The scalar fields undergo a two-step phase transition from the completely symmetric phase (white) to the intermediate phase (red), where only the singlet has a VEV, to the electroweak vacuum (yellow), where both Higgs and the singlet have VEVs. \textit{Right panel:} The two phase transitions. Evolution of the VEVs of the Higgs boson (green) and the complex singlet (red) with temperature. Critical temperatures are marked by thin vertical lines.}
\label{fig:phase:diagram}
\end{center}
\end{figure*}

\section{The Model}
\label{sec:model}

We consider the scalar potential of the SM Higgs boson $H$ together with a complex singlet $S$,
\begin{equation}
\begin{split}
  V &= \frac{1}{2} \mu_{H}^{2} \abs{H}^{2} + \frac{1}{2} \mu_{S}^{2} \abs{S}^{2} 
  + \frac{1}{4} \mu_{S}^{\prime 2} (S^{2} + S^{* 2})
  \\
  &  + \frac{1}{2} \lambda_{H} \abs{H}^{4} 
  + \lambda_{HS} \abs{H}^{2} \abs{S}^{2} + \frac{1}{2} \lambda_{S} \abs{S}^{4},
\end{split}
\end{equation}
where the $\mu_{S}^{\prime 2}$ term is the only one that softly breaks the $U(1)$ symmetry $S \to e^{i \alpha} S$. Without loss of generality, the parameter $\mu_{S}^{\prime 2}$ can be taken to be real and positive.

We decompose the fields in the electroweak vacuum as
\begin{equation}
  S = \frac{v_{s} + s + i \chi}{\sqrt{2}}, \quad 
  H = 
    \begin{pmatrix}
      0 \\ \frac{v_{h} + h}{\sqrt{2}}
    \end{pmatrix}.
\end{equation}
Note that both the Higgs boson and the singlet will get a vacuum expectation value (VEV) (the Higgs VEV is $v_{h} = 246.22~\mathrm{GeV}$). The mixing of the CP-even states $h$ and $s$ will yield two CP-even mass eigenstates $h_{1}$ and $h_{2}$. We identify $h_{1}$ with the SM Higgs boson with mass 
$m_{1} = 125.09~\mathrm{GeV}$ \cite{Aad:2015zhl}. Notice that the pseudo-Goldstone $\chi$ is the DM candidate with a mass determined by $ \mu_{S}^{\prime 2}.$

We express the potential parameters in terms of physical quantities in the zero-temperature vacuum, such as the masses $m_{1,2}^{2}$ of real scalars, their mixing angle $\theta$, pseudoscalar mass $m_{\chi}^{2}$, and the VEVs $v_{h}$ and $v_{s}$:
\begin{align}
  \lambda_{H} &= \frac{ m_{1}^{2} + m_{2}^{2} + (m_{1}^{2} - m_{2}^{2}) \cos 2 \theta }{2 v_{h}^{2}},
  \\
  \lambda_{S} &= \frac{m_{1}^{2} + m_{2}^{2} + (m_{2}^{2} - m_{1}^{2}) \cos 2 \theta}{2 v_{s}^{2}},
  \\
  \lambda_{HS} &= \frac{(m_{1}^{2} - m_{2}^{2}) \sin 2 \theta}{2 v_{s} v_{h}},
  \\
  \mu_{H}^{2} &= -\frac{1}{2} (m_{1}^{2} + m_{2}^{2}) + \frac{1}{2 v_{h}} (m_{2}^{2} - m_{1}^{2})
  \notag
  \\
  & \times (v_{h} \cos 2 \theta + v_{s} \sin 2 \theta),
  \\
  \mu_{S}^{2} &= -\frac{1}{2} (m_{1}^{2} + m_{2}^{2}) + 2 m_{\chi}^{2} + \frac{1}{2 v_{s}} 
  (m_{1}^{2} - m_{2}^{2}) 
  \notag
  \\
  & \times (v_{s} \cos 2 \theta - v_{h} \sin 2 \theta),
  \\
  \mu_{S}^{\prime 2} &= -m_{\chi}^{2}.
\end{align}

The tree-level direct detection DM amplitude vanishes at zero momentum transfer,
\begin{equation}
  \mathcal{A}_{\text{dd}}(t) \propto \sin \theta \cos \theta \left( \frac{m_{2}^{2}}{t - m_{2}^{2}} - \frac{m_{1}^{2}}{t - m_{1}^{2}} \right) \simeq 0,
\label{eq:cancel}
\end{equation}
which allows one to explain the negative experimental results from DM direct detection experiments, while
still keeping the pseudo-Goldstone DM mass in the reach of collider searches.

\section{Phase Transition}
\label{sec:phase}

In the high temperature limit, the $U(1)$-symmetric mass terms take on temperature-dependent corrections:
\begin{equation}
\begin{split}
  \mu_{H}^{2}(T) &= \mu_{H}^{2}(0) + c_{H} T^{2},
  \\
  \mu_{S}^{2}(T) &= \mu_{S}^{2}(0) + c_{S} T^{2},
\end{split}
\end{equation}
where 
\begin{align}
  c_{H} &= \frac{1}{48} (9 g^2 + 3 g^{\prime 2} + 12 y_{t}^2 + 12 \lambda_{H} + 4 \lambda_{HS}),
  \notag
  \\
  c_{S} &= \frac{1}{6} (\lambda_{S} + \lambda_{HS}).
\end{align}
The thermal correction to $\mu_{S}^{\prime 2}$ is zero, because the quartic couplings do not break the $U(1)$ symmetry.

The cancellation mechanism \eqref{eq:cancel} works only if the CP-even scalar states mix with each other. The mass matrix is non-diagonal only if both $h$ and $s$ get VEVs. For that, the fields must end up in the $(v_{h}, v_{s}, 0)$ vacuum at zero temperature. Note that we use e.g. $v_{h}$ as a label to indicate a non-zero VEV of the Higgs boson, not as a particular solution in terms of the potential parameters. Then the phase transition pattern consistent with the DM relic density is
\begin{equation}
  (0, 0, 0) \to (0, v_{s}, 0) \to (v_{h}, v_{s}, 0).
\label{eq:PT:pattern}
\end{equation}
Both steps are second-order phase transitions. 

There is no possibility to engineer a first-order phase transition. The only alternative second step, which could potentially be first-order~\cite{Vieu:2018nfq}, would be
\begin{equation}
 (0, 0, v_{\chi}) \to (v_{h}, v_{s}, 0).
\end{equation}
For a first-order phase transition, however, both extrema must be minima at the same time. But if the $ (v_{h}, v_{s}, 0)$ vacuum is a minimum, the $(0, 0, v_{\chi})$ vacuum can only be a saddle point or maximum, because the mass squared of the $s$ particle is $\mu_{S}^{\prime 2} < 0$ in this vacuum. When the potential contains a cubic term \cite{Jiang:2015cwa,Alves:2018jsw}, then the phase transition \eqref{eq:PT:pattern} into $(v_{h}, v_{s}, 0)$ can be of the first order, but such a term explicitly breaks the $\mathbb{Z}_{2}$ symmetry.

The phase diagram for one particular point of the parameter space with correct relic density \cite{Gross:2017dan} with the mixing angle $\sin \theta = 0.1$, the ratio $v_{h}/v_{s} = 0.291$, and masses $m_{2} = 1000~\mathrm{GeV}$ and $m_{\chi} = 100~\mathrm{GeV}$ is shown in Fig.~\ref{fig:phase:diagram}. In general, the allowed range of dark matter is between $60$~GeV and $10$~TeV~\cite{Gross:2017dan}, while in~\cite{Chiang:2017nmu} only a narrow range around the Higgs resonance was studied in detail as a viable parameter space for DM. The phase diagram in the left panel shows the evolution of fields (black line) from the $(0,0,0)$ vacuum (white) through the $(0, v_{s}, 0)$ vacuum (red) to the $(v_{h}, v_{s}, 0)$ vacuum (yellow). The phase where only the Higgs has a VEV is shown in green. The right panel demonstrates the two phase transitions. Both phase transitions are of the second order: in the first one the singlet VEV, in the second one, the Higgs VEV begins to grow continuously at the critical temperature, marked by the thin vertical line.

\vspace{0.5cm}
\section{Conclusions}
\label{sec:concl}

Pseudo-Goldstone DM in singlet scalar models with softly broken $U(1)$ presents an appealing possibility to sidestep constraints from direct detection on more general 
class of scalar singlet DM with a $\mathbb{Z}_{2}$ symmetry. Motivated by the aim to find additional tests of this framework we study the thermal phase transition pattern of the model. In order the model to work, the mechanism that cancels the direct detection cross section needs both the Higgs boson and the singlet to have VEVs.
For that reason, the possible phase transitions in this model are necessarily of the second order and, therefore, cannot produce any detectable gravitational wave signal.

Thus, a possible future discovery of a stochastic gravitational wave background characteristic to strong first-order phase transition would strongly disfavor or even rule out this class of models.
In this case the suppression of DM scattering cross section off nuclei must be explained by other means.

However, not discovering a signal would not rule out models with first-order phase transitions, because to generate a discoverable signal, the phase transition must be strong.

\begin{acknowledgments}

We would like to thank V. Vaskonen and A. Beniwal for discussions. This work was supported by the Estonian Research Council grant PRG434, the grant IUT23-6 of the Estonian Ministry of Education and Research, and by the EU through the ERDF CoE program project TK133.

\end{acknowledgments}


\bibliography{U1SBcxSM}

\end{document}